# Between Promise and Performance: Science and Technology Policy Implementation through Network Governance

*Science and Public Policy*, forthcoming

Travis A. Whetsell
Assistant Professor, Department of Public Policy & Administration
Florida International University
Travis.whetsell@fiu.edu

Michael J. Leiblein
Associate Professor, Fisher College of Business
Ohio State University
Leiblein.1@osu.edu

Caroline S. Wagner
Associate Professor, John Glenn College of Public Affairs
Ohio State University
Wagner.911@osu.edu

**Abstract:** This research analyzes the effects of U.S. science and technology policy on the technological performance of organizations in a global strategic alliance network. During the mid-1980s the U.S. semiconductor industry appeared to be collapsing. Industry leaders and policymakers moved to support and protect U.S. firms by creating a program called Sematech. While many scholars regard Sematech as a success, how the program succeeded remains unclear. This study re-contextualizes Sematech as a network administrative organization which lowered cooperation costs and enhanced resource combination for innovation at the cutting edge. This study combines network analysis and longitudinal regression techniques to test the effects of public policy on organizational network position and technological performance in an unbalanced panel of semiconductor firms between 1986 and 2001. This research suggests governments might achieve policy through inter-organizational innovations aimed at the development and administration of robust governance networks.

**Acknowledgements:** We are thankful for comments from Michael Siciliano, Noshir Contractor, Trevor L. Brown, Anand Desai, Allan Rosenbaum, Philip Russo, and anonymous reviewers. Data for this article were originally collected with support by the National Science Foundation under Grant #1133043.

# 1. Introduction

Government programs for science and technology (S&T) operate under conditions of uncertainty and complexity different from other kinds of governance frameworks (Hall and Lerner, 2010). Many studies on S&T policy rely on theory developed from research programs in economics (Malerba, 2002; Nelson and Winter, 1982), business and organizational studies (Teece, Pisano, and Shuen, 1997), and patenting activities (Hall and Ziedonis, 2001). Work in this area often lacks an emphasis from public policy and administration. Bringing this perspective to the fore, this article synthesizes disparate literatures, combining elements from network governance (e.g. Provan and Kenis, 2008; Klijn and Koopenjan, 2015) with elements from organization theory and strategic management (e.g. Eisenhardt and Schoonhoven, 1996; Stuart 2000). Insights from these literatures are leveraged to analyze an exemplar case of government intervention in the high-technology sector.

This study analyzes the implementation of U.S. technology policy during the early evolution of a cooperative research and development (R&D) network in the semiconductor industry between 1986 and 2001. The focus is on a U.S. Department of Defense (DOD) sponsored consortium, called Sematech, its hypothesized effects on an emerging R&D alliance network, and hypothesized effects on technological performance. The Sematech consortium was formed to provide support for the U.S. based semiconductor industry during the mid-1980s (Macher, Mowery, and Hodges 1998; Browning and Shetler 2000). Many scholars suggest the policy experiment was a success, as the downward trend for U.S. semiconductor market share was reversed and the industry remains among the top five U.S. exports with sales valued at roughly $165 billion, capturing 50% of the global market in 2015 (SIA Databook, 2016).

However, the case presents an interesting theoretical puzzle. A view from public administration might suggest that public support for the consortium lowered R&D costs for private firms, overlooking the relevance of synergies created through an evolving network of relationships. Conversely, a business view might suggest that strategic alliances enhance access to resources for competitive advantage, downplaying government support, protection from foreign participants, and ignoring a major political crisis with national security overtones. Putting the puzzle together requires a pluralistic integration of disparate disciplines and theories (Whetsell 2013). The research question of this article is, *how* might a network administrative organization (NAO), such as Sematech, achieve effectiveness given a range of potential theoretical explanations? An overlooked element is the intermediate network mechanism residing between S&T policy and performance, where we conceptualize the network position of an organization as the intermediate mechanism. An empirical analysis of this mechanism, as well as its linkages to policy and performance, provides insights into the specific case but also has more general implications for government investment in science and technology.

The theoretical argument of this article is that cooperative governance structures may be useful for addressing market failures on volatile technological landscapes. The more specific theoretical logic suggest that the Sematech consortium represents a mode of network governance (e.g. Klijn and Koopenjan 2015), structured as a network administrative organization (e.g. Provan and Kenis 2008), which decreased the costs of cooperation (e.g. Oxley, 1997), improved access to complementary resources for innovation (e.g. Barney 1991; Eisenhardt and Schoonhoven, 1996), and enhanced the social capital and cooperative capacities of participating organizations for competitive advantage (e.g. Ahuja 2000; Stuart 2000; Zaheer and Bell 2005). The empirical approach of this study is to test hypothesized effects of Sematech membership on the network position of member firms, as well as test hypothesized effects of enhanced network position on firm level technological innovation. The basic model is that the performance effects of Sematech flow through network position. In addition, this research also tests for spillover effects on strategic allies and potential effects of DOD exit from the consortium.

The following section briefly describes the historical context of the case necessary to establish a timeline of events. Section three synthesizes a set of theories from the public and private sectors to suggest hypotheses. Section four presents the data collection and analysis methods. Section five shows the results of the analysis. Section six discusses the implications of the results. Finally, we close with some concluding remarks about the subject and the study.

## 2. Historical Context & Case Details

Government investment in science and technology (S&T) has been a central national priority for public policy makers and administrators in the United States since at least World War II (Bush, 1945). In economic terms, S&T are curious sorts of goods. In contrast to basic goods and services, the uncertain and risky nature of S&T activities, as well as the difficulty of appropriating returns on research and development (R&D), means they tend to suffer from underinvestment in private markets, despite the more general contribution to economic growth (Arrow, 1962; Partha and David, 1994; Stephan, 1996). Science is often defined as a public good, where other actors cannot be excluded from appropriating its value, and many actors can make productive use of scientific knowledge simultaneously (Samuelson, 1954; Stiglitz, 1999). These insights suggest that pure markets do not provide the necessary resources for the conduct of basic science (Nelson, 1959). In contrast to science, technology has characteristics that make it excludable in many cases (Rosenberg, 1982). Technological processes and products may be patented and sold as private goods. Thus, private returns of technology investment may be appropriated by the investor. Nevertheless, pure markets for technology also tend to suffer poor incentives, high uncertainty, and weaker appropriability of value than basic consumer goods (Arrow, 1962; Stephan, 1996). The economic logic of S&T market failure has been a part of the rationale for government action since early 20$^{th}$ century. After World War II a mission-based logic emerged emphasizing the role of S&T in national defense and other policy or agency specific objectives. Toward the end of the 20$^{th}$ century a cooperative boundary spanning logic emerged which supported industrial policy and regional economic development (Bozeman 2000; Salter and Martin, 2001).

In the 1980s, the pace of technological change began to challenge older models, undermining the neoclassical distinction between public and private sectors (Smith, 1990), as well as the assumption that technology is exogenous to the production function (Romer, 1990). Exponential increases in technological performance and manufacturing costs accompanying the emergence of the microprocessor led to major changes in the competitive behavior of firms in tech-based sectors. The challenge of maintaining innovation produced significant increases in inter-firm cooperation. Strategic alliances emerged in the 1980s across several sectors. Research documenting increases in R&D-based alliances during this period suggest that private firms were increasingly collaborating to reduce risk and to share costs (Hagedoorn, 2002). Further, alterations in federal anti-trust policy, e.g. the 1984 National Cooperative Research Act, provided a more permissive atmosphere for R&D collaboration between private sector firms (Mowery 1998).

In 1985 Japanese firms took the majority global market share of semiconductor sales. This shocking development was an achievement of the Japanese organizational conglomerate system, known as the *keiretsu*, which resembled a networked form of organization, designed to coordinate increasingly costly R&D activities and facilitate quicker development of strategic technologies. Japanese firms also received strong support from major government R&D programs (Ham, Linden, and Appleyard,1998; Sakakibara, 1993). This development suggested that the Japanese organizational form may have been more resilient and better adapted to absorbing the risks associated with the new global environment for technological growth.

In response, American leaders in the semiconductor industry and Congress crafted an innovative policy approach to address the imminent failure of the U.S. based industry. An alignment between economic and defense interests facilitated consensus on the policy response. The economic interests emphasized preservation of a top U.S. export, supporting high-quality jobs, and feeding into numerous other products, i.e. a "platform technology". The defense interests characterized semiconductors and integrated circuits as a critical resource necessary to maintain the cutting edge in high-technology weapon systems (Mowery 1983; Mowery and Langlois 1996; Mowery 2009).[1] Rather than relying on subsidizing firms directly or imposing tariffs and trade barriers, industry leaders and policy makers crafted an organizationally-based policy solution.

The result was the non-profit public-private consortium, <u>Se</u>miconductor <u>Ma</u>nufacturing <u>Tech</u>nology, referred to here as "Sematech", created in 1987 with support from the Department of Defense (DOD) and the Defense Advanced Research Projects Agency (DARPA). The Sematech consortium can be characterized as a type of strategic partnership or alliance (Siegel and Zervos 2002). The consortium began as a multilateral agreement between DOD and fourteen U.S. semiconductor manufacturing firms, constituting roughly 85% of U.S. manufacturing capacity. The consortium received roughly $100 million per year for two five-year periods from DOD. Sematech members contributed a minimum of one-million dollars, or one percent of sales, with a fifteen-million-dollar cap. Sematech represents a case of -self-organization with leadership by the members themselves, and a relatively hands-off role of DOD (Beyer and Browning 1999). Members were also required to contribute personnel to joint R&D activities at the Fab One facility in Austin, TX, where they engaged in a collocated, face-to-face effort to conduct "pre-competitive" R&D. The results were shared among the members and applied for competitive advantage relative to foreign firms, primarily in Japan. The consortium excluded foreign firms from participating from 1987-1995. However, there were no restrictions on consortium members forming alliances with foreign firms outside of the Sematech consortium (Browning & Shetler 2000). DOD sponsorship and the prohibition on foreign participation ended in 1996, but Sematech continued until the present as a non-profit consortium.

---

[1] The interplay of economic and defense interests can be observed in a 1989 Senate Armed Services Subcommittee on Defense Industry and Technology hearing on The Future of The Semiconductor Industry -- https://www.c-span.org/video/?10092-1/future-semiconductor-industry

**Figure 1 – Timeline of Significant Events**

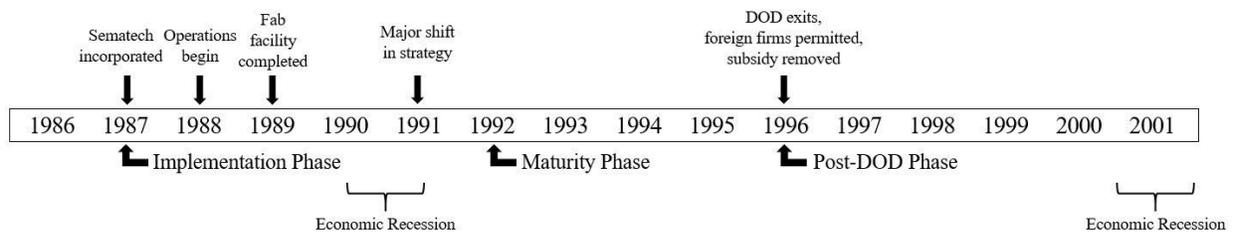

**Figure Notes** – The figure shows the timeline of significant events, which is divided roughly three periods, implementation/DOD, maturity/DOD, and post-DOD period. Two economic recessions occurred during the study period. The maturity phase occurs after a major shift from a focus on horizontal to vertical relationships in the consortium.

Figure 1 shows a timeline of events, which is demarcated into an implementation period, a maturity period, and a post-DOD period. This timeline was established based on insights from the extant qualitative literature on Sematech (Grindley, Mowery, and Silverman, 1994; Browning and Shetler, 2000; Carayannis and Alexander 2004; Bonvillian 2013). This timeline is important to the subsequent analysis.

## 3. Literature Review

*Government & Network Effectiveness*. The literature on network governance provides useful concepts for thinking about the processes and outcomes of cross-sector policy implementation regarding inter-organizational networks (Provan and Kenis, 2008; Klijn and Koopenjan 2015). However, scholars have not sufficiently applied these policy frameworks to science and technology (S&T) policy.[2] As Laranja (2012) suggested, network governance lacks a comprehensive theoretical application to S&T innovation. More generally in public policy and administration studies, Berry et al. (2004) suggested there is a lack of attention to network evolution, while Provan, Fish, and Sydow (2007) suggested that more attention needs to be paid to the dynamics of networks over time and the effect of networks on outcomes. This article addresses these issues by applying the network governance framework to analyze an exemplar case of S&T policy in terms of both temporal dynamics and organizational performance.

Klijn and Koopenjan (2015:11) define governance networks as "stable patterns of social relations between mutually dependent actors, which cluster around a policy problem, a policy programme, and/or a set of resources and which emerge, are sustained, and are changed through a series of interactions". In this study, the imminent collapse of a U.S. industry is the policy problem, the policy program is the Sematech consortium, and the spread of cooperative partnerships across the industry represents an emerging pattern of social relations, that permits the combination of heterogeneously distributed resources, broadly defined as tangible or intangible from physical capital to information (e.g. Barney 1991).

---
[2] Notable exceptions include Kash & Rycroft (1998) and Wardenaar, de Jong, and Hessels (2014).

From this broad definition of governance networks follows a more specific conceptual device, called a network administrative organization (NAO), which helps to further recontextualize the case. Provan and Kenis (2008:236) define NAOs as separate administrative entities, usually government or non-profit organizations, which serve as governance mechanisms that coordinate and sustain the network of interactions between organizations, whether public, private, or non-profit. The network governance treatment of Sematech as NAO shifts the emphasis away from economic policy instruments, such as direct subsidies and tariffs, to an emphasis on the complex interactions between an evolving network of organizations. As Kenis and Provan (2009) suggest, NAOs are often adopted when establishing legitimacy and building sustainability are particularly important. Human and Provan's (2000) analysis of small and medium sized enterprises illustrates the need for establishing legitimacy in private sector networks where the norm is competition rather than cooperation. The NAO helps to build the kind of legitimacy necessary for aligning organizational objectives with collective goals. The extant literature on Sematech reveals the difficulties of aligning interests during the early implementation of the consortium, particularly around secrecy and information sharing (e.g. Browning, Beyer, & Shetler 1995; Browning & Shetler 2000). Insights from network governance help to extend our understanding of Sematech, where the consortium members were also competitors in a fierce environment of rapid technological innovation.

The NAO model contrasts with the lead organization model and participant governed model of network governance (Provan and Kenis 2008), which are less applicable to the current case. However, Wadrenaar, de Jong, and Hessels (2014) suggest this typology of network governance is a suitable framework for understanding various types of strategic research consortia in the S&T setting. Applying these insights to the current case, the lead organization model might apply if, for example, the Department of Defense had taken a lead role in the semiconductor industry network. Conversely, the participant-governed model has no lead organization or distinct administrative entity set up to govern activities and is a decentralized model of interaction. Both alternatives to the NAO model lack the network capacity necessary to govern interactions between a large number of firms. The lead organization model is too centralized and inflexible given the self-organizing nature of technological innovation, while the participant governed model is too decentralized and unstable to effectively steer the U.S. high-tech sector during an existential crisis. The question is, what governance form is appropriate to the specific context in order to achieve network effectiveness?

Provan and Kenis (2008:230) define network effectiveness as the achievement of collective outcomes not otherwise possible by individual organizations alone. Yet specification of effectiveness varies across different contexts (Kenis and Provan 2009), particularly within cross-sector networks where organizations may have distinct goals (Raab, Mannak, and Cambre 2013; Klijn and Koopenjan 2015). In the case of Sematech, the goals were relatively clear cut: increasing process quality, and innovation in semiconductor architecture miniaturization (Browning & Shetler 2000; Epicoco 2013). Sematech's "Black Book" set roadmap milestones according to the progressive miniaturization of integrated circuit feature line sizes at the micron scale (Browning & Shetler 2000). Given these insights, we define network effectiveness as the ability of Sematech members to achieve superior innovation, relative to non-members, according to miniaturization of semiconductor line widths.

However, there are limitations regarding the conceptualization of government intervention in private sector networks. While the networks literature in public administration has examined the roles of interaction, collaboration, trust, and reciprocity in multiple settings (e.g. Agranoff and McGuire, 2001; O'Toole and Meier, 2004; Ansell and Gash 2008; Emerson, Nabatchi, and Balogh, 2012), the focus is often on directive relationships between principals and agents (Provan and Milward, 1995), resource-dependencies (Rethemeyer and Hatmaker, 2008), and power dynamics (Saz-Carranza, Iborra, Albareda 2016) in the public and non-profit sectors. This approach can be limiting in a private sector context, since conceptualizing network relationships in terms of power-based dependencies can fail to explain the strategic decision-making processes of private firms. As we suggest, it may be more useful define the role of government as a potential catalyst for cooperation between competitors on scientific and technological landscapes (Kash and Rycroft 1998).

Finally, while the network governance literature provides important descriptive concepts (cf. Provan & Lemaire 2012), it does not necessarily motivate specific hypotheses about the effect of government on network dynamics or organizational performance in the S&T context. The following section integrates this framework with theories necessary for articulating more localized hypotheses about how network governance might be used to enhance cooperation and innovation in the high-technology sector.

*Strategic Alliances & Network Position.* The literature on strategic alliances illustrates the necessity of deploying multiple theories to explain cooperative behavior between organizations. Principally, these are transaction cost economics (TCE), the resource-based view of the firm (RBV), and social network theories, e.g. social capital theory. The application of transaction cost economics provides a robust explanatory framework for strategic alliance behavior in the semiconductor industry (e.g. Oxley, 1997; Yasuda, 2005; Leiblein and Macher, 2009). With the rapidly rising costs of R&D essential to semiconductor manufacturing, vertically integrated but isolated firms in pure competition could no longer maintain progress along the semiconductor miniaturization trajectory of Moore's Law (Epicoco, 2013).[3] Strategic alliances emerged to facilitate the kind of close cooperation necessary for continual innovation. However, cooperation is not frictionless. As Williamson (1981,1985,1991) argued, the transaction costs of cooperation emanate primarily from the uncertainty and opportunism associated with human nature. These factors are particularly salient when cooperative transactions are uncertain, recur frequently over time, and assets are transaction specific, all of which characterize cooperative R&D for high-technology. To reduce transaction costs, cooperative alliances in the semiconductor industry are often governed by a mix of bilateral and multilateral cross-licensing, co-development, and joint venture agreements.

Applying the logic of transaction costs, Oxley (1997) reasoned that transaction costs increase with each additional R&D alliance partner, showing that alliance scope is associated with increasingly hierarchical forms of governance from simple licensing agreements to joint ventures. Multi-firm alliances are common in the semiconductor industry because innovation often requires the application of resources by multiple firms, and greater efficiency may be achieved through a single multilateral governance structure rather than through a series of bilateral agreements. Multilateral alliances may also propagate effects on the broader network of relationships (Persidis and Persidis, 1996; Medcof, 1997; Hwang and Burgers, 1997). In terms of network theory, multi-firm alliances could be treated as cliques, where all possible pathways between partners are realized (Knoke and Yang 2008); and, each addition to an alliance clique greatly increases the number of connections in the alliance. This suggests a greater capacity for sharing and combination of resources in multi-firm alliances (Hage, Jordan, & Mote 2007). Such configurations may also facilitate the emergence of network safeguarding mechanisms that serve to protect the pattern of exchanges in the network, such as restricting access, imposing sanctions, and reputation management (e.g. Jones, Hesterly, Borgatti 1997). We conceptualize the Sematech consortium as a type multi-firm alliance, roughly double the size of the largest extant alliance and, therefore, permitting resource combination to a higher degree.

---

[3] Moore's Law describes the exponential miniaturization of semiconductor devices, where the number of transistors that can be placed on an integrated circuit doubles roughly every two years (Epicoco, 2013:181).

However, large multi-firm alliances come with even greater costs (Li et al., 2012; Gudmundsson, Lechner, and Van Kranenburg, 2013). At a certain point the costs of additional alliance partners may overwhelm the benefits of resource combination. When the objectives of such alliances are public in nature or when governments view them as necessary to fulfill public objectives, publicly supported governance regimes such as network administrative organizations (NAOs) might provide the supportive structure necessary for managing the costs of cooperation. Thus, we integrate TCE with the network governance literature by suggesting that NAOs reduce the transaction costs associated with large-scale cooperation.[4]

Finally, synthesizing these insights with social network theory, we advance the logic for how network governance enhances firm level network position. In social network analysis, centrality characterizes the network position of an actor representing prominence, popularity, or power within a network, such that actors high in centrality are "well-connected" to other actors in the network (Wasserman and Faust 2007; Hanneman and Riddle 2011; Scott, 2017). The most basic centrality measure is degree centrality, which simply counts an actor's ties. We use eigenvector centrality, which represents how well-connected an actor is to other well-connected actors. Theory suggests that being connected to other well-connected organizations enhances access to complementary resources necessary for innovation at the cutting edge. We suggest that NAOs reduce the cooperation costs of strategic alliances leading to increases in cooperative activity, which entails an increase in network centrality relative to non-NAO members. Thus, we hypothesize that Sematech enhanced the network centrality of member organizations by lowering the costs associated with cooperation on pre-competitive research and development.

---

[4] Williamson (1999) makes a very similar argument about the necessity of government to maintain transaction costs between nations embodied in the U.S. Department of State.

> *H1: Network Administrative Organizations enhance the network centrality of member organizations relative to other organizations within emerging strategic alliance networks.*

The resource-based view (RBV) suggests that firms gain an advantage by applying resources in a competitive environment (Wernerfelt, 1984). Barney (1991:101) defines resources as "all assets, capabilities, organizational processes, firm attributes, information, knowledge, etc. controlled by a firm that enable the firm to conceive of and implement strategies that improve its efficiency and effectiveness". The RBV highlights the importance of difficult or impossible to trade (often intangible) resources in explaining performance heterogeneity and persistence (e.g., Barney, 1991; Dierickx and Cool, 1989). The RBV also calls attention to information and knowledge as a resource feeding into organizational performance (Grant and Baden-Fuller, 2004). In the high-technology sector, information, knowledge, and intellectual property are among the most critical resources for competitive advantage.

Extensions of the RBV through social capital theory suggests that social relationships provide access to heterogeneously distributed resources (Eisenhardt and Schoonhoven, 1996;

Tsai and Ghoshal, 1998; Nahapiet and Ghoshal, 1998). Social capital theory focuses attention on the ways in which firms seek to leverage their relationships for competitive gain (1997; Lin 1999). Provan & Lemaire (2012) suggest that organizations gain the advantages of social capital when they form network ties. As Walker, Kogut, and Shan (1997) suggest, the "resource view" of social capital is concerned with the advantages of relationships for individual firms, rather than emphasizing group level processes, such as influence, status, and prestige (e.g. Putnam, 1995; Podolney, Stuart, Hannan 1996). Similarly, scholars have distinguished between the structural, cognitive, and relational dimensions of social capital (Nahapiet and Ghoshal, 1998; Tsai and Ghoshal 1998; Inkpen and Tsang, 2005). For the purposes of this article, we take a structural approach, emphasizing a network- and resource-based conceptualization of social capital (Lin 1999; Burt 2000). Lin (1999:7) defines social capital as "resources embedded in a social structure which are accessed and/or mobilized in purposive actions". This concept suggests why the technological innovation of the firm depends to some extent on its positioning within a strategic alliance network.

Previous research has examined the effects of alliance formation on performance and innovation in the semiconductor industry (Eisenhardt and Schoonhoven, 1996; Stuart 2000; Hill, Jones, and Schilling, 2014; Schilling 2015), as well as the effects of firm positioning within alliance networks on performance (Ahuja 2000). Studies have examined the effects of network centrality on organizational performance (Zaheer & Bell 2005; Koka & Prescott 2008). Social capital extensions of the resource-based view suggest that network centrality measures a firm's access to heterogenous and immobile[5] resources in the network (Walker, Kogut, & Shan 1997). Leveraging these insights, we advance the following hypothesis.

---

[5] Resource immobility refers to the inability to trade resources between organizations (Peteraf 1993). They are often a source of competitive advantage, and firms tend to form alliances in order gain access to these resources.

*H2: Network centrality enhances the technological performance of organizations in strategic alliance networks.*

Connecting the logic of H1 and H2 entails that Sematech membership enhances the technological performance of member firms through enhancements in network position. This logic suggests a mediation model (e.g. Baron & Kenny 1986; Aguinis et al. 2017). Thus, we hypothesize that the direct performance effects of Sematech membership flow through its effects on the network position of member firms.

*H3: Network centrality mediates the effects of Network Administrative Organization membership on technological performance relative to other organizations in strategic alliance networks.*

Since network processes are relational, we suspect that strategic allies of Sematech members will experience similar enhancements in network position and technological performance relative to firms not directly connected to Sematech members. This is similar to the logic of policy diffusion (e.g. Shipan & Volden 2012), but the focus here is on the policy effects on technological diffusion through network spillovers. The spillover hypothesis was advanced by previous studies (Irwin and Klenow, 1994) but has never been directly applied to this case through network analysis. Here, Sematech is thought to have secondary social capital effects, or spillovers, on allies of Sematech members.

*H4: The network and performance effects of Network Administrative Organization membership spillover onto strategic allies of NAO members relative to other organizations in alliance networks.*

Finally, the timeline of events displayed in Figure 1 suggests that the development of Sematech within the strategic alliance network moved through three periods during the study period: implementation, maturity, and post-DOD. Since technological performance data are not available for the implementation period, the maturity period is compared to the post-DOD period. Since Sematech continued to exist as a non-profit after DOD exit, quasi-experimentally, this amounts to a partial removed treatment effect of DOD exit. DOD exit in 1996 ended the subsidy for the consortium and permitted foreign firm membership in Sematech. While the removal of the matching subsidy may have reduced Sematech network capacity, the simultaneous entry of new powerful foreign firms may have had the opposite effect. Given the exploratory nature of the proposition, we advance the following working hypothesis[6] that the effects of Sematech on network position and technological performance, as well as mediation effects, may be stronger during the DOD sponsorship period.

*H5: NAO and network effects are stronger during the DOD sponsorship period than in the post-DOD period.*

---

[6] See Shields and Rangarajan (2013) for a discussion of working hypotheses in exploratory research.

**4. Methods**

*Data*. To construct the dataset for the study period (1986-2001), alliance data were gathered from two sources. First, from 1986-1989 alliance data were gathered from public announcements compiled through press releases and other public news announcements. Second, alliance data from 1990-2001 were extracted from the ICE/IC Insights Strategic Profiles Reports on the global semiconductor industry. The first data source reports only announcements, while the second also reports ongoing alliances. The early data between 1986 and 1989 likely undercounts alliances. Data on firm sales and technological performance were also gathered from these sources. Missing data was gathered from COMPUSTAT/CRSP via WRDS, Bloomberg, S&P Capital IQ, and PrivCo. Sematech membership data were acquired from a contact with Sematech. Since data on technological performance is only available starting in 1990, the sample frame for models using technological performance was reduced to 1990-2001.

The network data were constructed in the following manner. All alliances listed in the ICE/IC Insights profiles were aggregated yearly and constructed as symmetrical adjacency matrices, a common method for analyzing whole-networks over time (Wasserman and Faust, 1994). Each year-matrix is one-mode and undirected, with binary values that represent the presence or absence of an alliance edge between firm nodes. Isolates were not included. The edge weights for each year were set to one to handle cross-listing of alliances in the IC Insights profiles. After constructing the yearly adjacency matrices, they were analyzed, and node level network statistics were calculated using the network analysis program Gephi. Node level measures from each whole-network year were then extracted and merged with firm level panel data. Figure 2 shows the development of these networks across the time-period. In these visualizations. A conservative approach is taken in the networks, where ties between Sematech members are not included unless they are alliances external to the consortium.

**Figure 2 – Yearly Strategic Alliance Network 1986-2001**

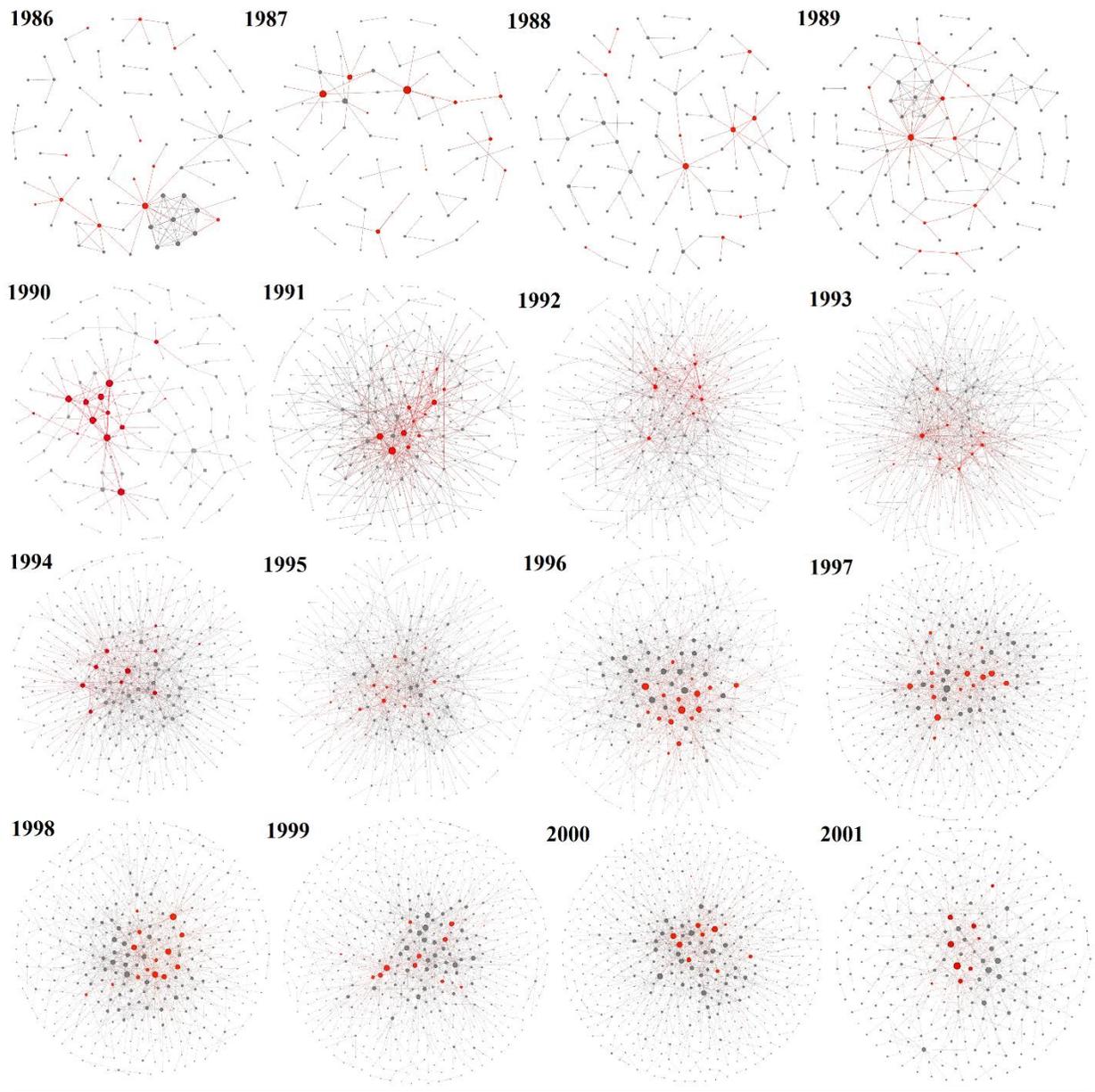

**Figure notes** –The networks are constructed as one mode, undirected adjacency matrices, where edges are alliances and nodes are firms. Red nodes are Sematech members, and red edges are connections with Sematech members. Red nodes in 1986 show the future Sematech members. Nodes are sized by degree centrality. Alliances from 1986-1989 are likely undercounted due to different method of collection.

***Sample*.** Sematech members include large and medium sized integrated circuit manufacturers. The original members included AMD, AT&T, Digital, Harris, HP, IBM, Intel, LSI Logic, Micron, Motorola, NCR Corp, National, Rockwell, and TI. Few changes in consortium membership occurred in the implementation phase; it lost members during the maturity phase; and the consortium increased in size as foreign firms joined in 1996. Despite being prominent firms, the collective performance crisis for U.S. firms (Hof 2011) may temper bias from initial selection into Sematech and subsequent performance. A small number of Sematech members were not included in the analysis if they also did not appear in the alliance data from IC Insights.

The larger sample of semiconductor firms included in the broader alliance network were also captured through the IC Insights Profiles. These include a heterogenous group of large, medium, and small firms, as well as manufacturers, suppliers, and pure-play IP foundries. Since firms with no recorded alliances are not included, the analysis is limited only to those firms with alliances. As with most non-experimental data, there are missing data in our sample. Thus, the number of observations varies across our tables presenting descriptive statistics, correlations, and regression models. The different sample sizes are the result of variance in the frequency of missing data across variables and across temporal panels. Thus, the final sample used in the models has an unbalanced panel structure, which include between 150 to 300 firms depending on the model and time period specification.

***Measures*.** Eigenvector centrality (E.Centrality) was chosen as the variable characterizing organizational network position (see Newman 2018:159). In the present study, eigenvector centrality represents how well-connected a firm is to other well-connected firms in the network (Borgatti, Jones, and Everett 1998; Zaheer, Gözübüyük, and Milanov, 2010). Eigenvector centrality captures network position of a firm better than a basic measure like degree centrality, since it includes information not only about a firm's number of partners but also about the prominence of their partners (Newman 2018). Hence, this variable captures information about both direct and indirect ties and is among network measures known to impact firm performance (Uzzi, 1996; Zaheer and Bell, 2005). Eigenvector centrality was calculated for all firms in each network-year cross section then merged with the firm level data.

The variable for technological performance (Tech-Performance) is the minimum integrated circuit feature size that a firm can produce, given yearly data available on manufacturing facilities. This is a unique measure specific to the semiconductor industry, where the minimum integrated circuit feature size is an objective measure of technological sophistication. This variable has been used previously in studies on the semiconductor industry (Eisenhardt and Schoonhoven 1996; Leiblein, Reuer, and Dalsace 2002). Since the measure is defined in terms of a continuously decreasing minimum, this variable is reverse coded to be consistent with performance. However, the technological performance data have limited coverage, compared to the network centrality data. To understand how missing data for this variable affect the analysis, we conducted a t-test on the difference on size (annual sales), centrality, and headquarters location between firms with and without data on technological performance, indicating smaller, less central, US firms tended to have less coverage. The technological performance data is derived from data on manufacturing facilities. Many firms focus on supplies and intellectual property rather than manufacturing. Thus, the analysis of technological performance generally applies to manufacturing firms that are medium to large, with relatively higher network centrality, and with a diverse mix of nationalities.  The policy variables include Sematech and S.Partners. Sematech is a nominal variable, taking the value of 1 if the firm is a Sematech member at time *t*, and 0 otherwise. S.Partners is also a nominal variable taking the value of 1 if a firm has a strategic alliance with a Sematech member at time *t*, and 0 otherwise. The control variables include Org.Size and US. We operationalize firm size in terms of annual revenue or sales in millions. Measures such as employee count are likely to be contaminated by the decision to vertically integrate into production or assembly. More specifically, organizational size is operationalized as a three-year moving average of total sales. A moving average was used given the cyclical and volatile nature of sales in the semiconductor industry. Finally, US captures the nationality of the headquarters of the firm; the variable takes a value of 1 for US firms, and 0 otherwise.

*Methods*. The primary method of analysis employed is the mixed-effects model, which combines elements from fixed and random effects models. The mixed effects model is a generalization of the standard linear model, which allows for modeling the means and the variance/covariance of the data (Littell et al., 2006). Mixed models were chosen to account for correlation within repeat observations on firms without loss of meaning regarding fixed firm characteristics (*see* Allison, 2005), such as firm nationality, Sematech membership which only occasionally changes for some firms, and for technological performance which often remains identical for multi-year stretches of time. Since the modeling strategy is to compare mediation effects and models in two separate four-year periods (Table 2 & 3), there isn't enough variability to produce useful estimates in fixed effects models. However, robustness checks were conducted using firm and year fixed effects on the full time period (Table 4). In the mixed effects model, the unique firm identifier is the covariance parameter, which is modeled as a random effect for repeat observations over time. The variance components of the random parameter (firm id) include a covariance estimate and a residual. These two pieces of information are used to calculate the intraclass correlation (ICC), which estimates the amount of variability in the outcome accounted for by the firm identifier and the residual (Allison, 2005).

## 5. Results

Table 1 presents the descriptive statistics and bivariate correlations between all the variables used in the subsequent models. The bivariate correlations show significant relationships between Sematech members (Sematech), network position (E.Centrality), and technological performance (Tech-Performance), and a similar but weaker pattern is observed for allies of Sematech members (S.Partners). Additionally, E.Centrality has a significant positive association with Tech-Performance. The existence of these intercorrelations conform with expectations and suggest the need for multivariate analysis.

**Table 1 – Descriptive Statistics and Correlation Matrix**

| Variable | N | Mean | Std Dev | Min | Max | 1 | 2 | 3 | 4 | 5 | 6 |
|---|---|---|---|---|---|---|---|---|---|---|---|
| 1 - Sematech | 3200 | 0.058 | 0.23 | 0 | 1 | 1 | | | | | |
| 2 - S.Partners | 3200 | 0.154 | 0.36 | 0 | 1 | -0.106 | 1 | | | | |
| 3 - Tech-Performance | 966 | 59.317 | 29.54 | 1 | 104 | 0.259 | 0.086 | 1 | | | |
| 4 - E.Centrality | 1962 | 0.167 | 0.214 | 0 | 1 | 0.464 | 0.191 | 0.468 | 1 | | |
| 5 - Org.Size | 2581 | 592.539 | 1602 | 0.027 | 26260 | 0.465 | 0.001 | 0.372 | 0.534 | 1 | |
| 6 - US | 2986 | 0.687 | 0.464 | 0 | 1 | 0.076 | 0.003 | -0.155 | -0.029 | -0.100 | 1 |
| 7 - Year | 3200 | 1995 | 4.222 | 1986 | 2001 | -0.037 | 0.097 | 0.538 | 0.105 | 0.116 | -0.086 |

Table notes: The table shows the descriptive statistics and the correlation matrix of all variables used in the analysis. The sample sizes for each variable are different due to the uneven data coverage in the unbalanced panel data.

Figure 3 shows the comparison of mean values for eigenvector centrality across the study period for 1) Sematech members, 2) partners of Sematech members, and 3) non-members/non-partners. 1986 shows the mean value for future members in 1987 prior to Sematech formation. The figure shows that each group begins at similar levels of network centrality. Then, Sematech members and partners of Sematech members experience dramatic increases in centrality across the period, with reductions after DOD exits the consortium. It should be reiterated that 1986-1989 report only contemporary announcements and do not report ongoing alliances, likely undercounting alliances in the network during this period. Further, a conservative approach was taken for the full analysis in which ties between Sematech consortium members were not included in the networks; hence, productive alliance ties between Sematech members are also underestimated in these models.

**Figure 3 – Sematech & Eigenvector Centrality**

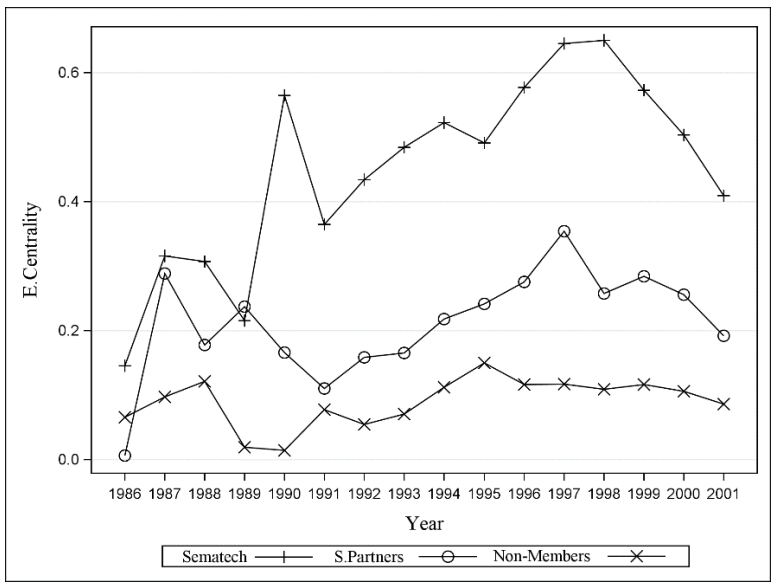

**Figure notes** - The figure shows the average eigenvector centrality for Sematech members (+), allies of Sematech members (o), and non-member/non-allies (x), yearly across a sixteen-year period. Sematech was formed in 1987. Thus, 1986 shows the centrality of future members in 1987. 1987-1991 is the implementation phase. 1992-1995 is the maturity phase. 1996 is the beginning of the post-DOD phase.

Figure 4 shows the comparison of mean values for technological performance across the study period for 1) Sematech members, 2) partners of Sematech members, and 3) non-members/non-partners. A similar trend to Figure 3 is observed. However, the trend appears much smoother. All firms experience an increase in technological performance across time; Sematech members have lower performance in 1990; but Sematech members and their allies experience greater gains.

**Figure 4 – Sematech and Technological Performance**

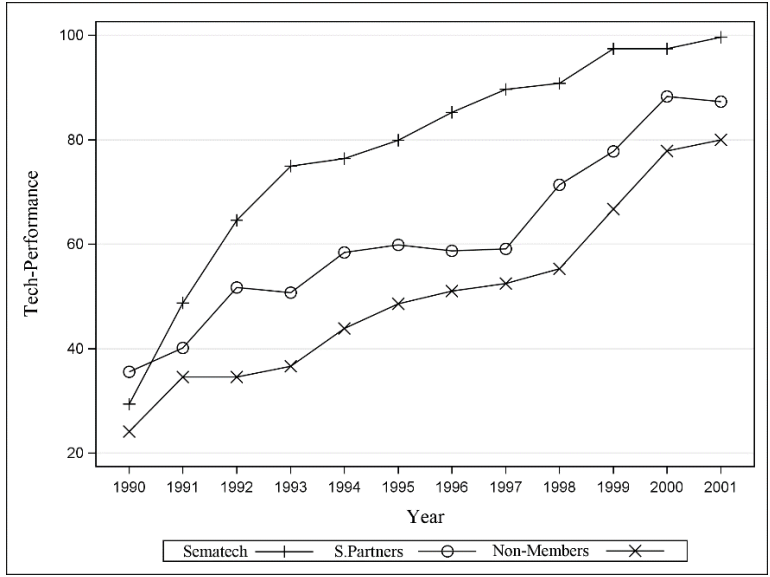

**Figure notes** - The figure shows the average technological performance for Sematech members (+), allies of Sematech members (o), and non-member/non-allies (x), yearly across a twelve-year period. 1992-1995 is the maturity phase. 1996 is the beginning of the post-DOD phase.

Hypothesis tests are presented in Table 2, which has eight mixed effects models, separated by DOD period and post-DOD period. As is illustrated in Figure 1, 1992-1995 represents the maturity period of Sematech, where the consortium receives DOD support and protection; after 1995 represents the post-DOD period, where support is removed, and foreign firms are permitted entry. These four-year windows were chosen to avoid impacts of two major economic recessions occurring around 1990 and 2000. Technological performance data was unavailable during the implementation period, between 1987 to 1990.

Model 1 in table 2 shows that Sematech and S.Partners have much larger estimates on E.Centrality than non-members, providing support for hypothesis one and hypothesis four.[7] Model 2 shows that Sematech members also have a larger estimate Tech-Performance, but S.Partners do not. Model 3 shows that the E.Centrality is significantly associated with Tech-Performance, providing support for hypothesis two. Model 4 shows that when including E.Centrality in the same model with Sematech, the estimate on Sematech is reduced from 19.452 to 10.039 and the estimate is no longer statistically significant (p, 0.12), providing support for mediation in hypothesis three. Following the mediation analysis procedure by Baron & Kenny (1986), the results of models 1-4 suggest E.Centrality acts as a mediator through which the performance benefits of Sematech flow. The Sobel Test also suggested the indirect effect of Sematech (T-stat, 4.17; p,0.00003) on tech-performance via network centrality is significantly different from zero (*see* Preacher & Leonardelli 2001).

Hypothesis four, regarding the spillover effects on partners of Sematech members, is supported in some models but not others. Being an ally of Sematech had a significant estimate on E.Centrality but not directly on Tech-Performance. However, the mediation hypothesis may apply to allies of Sematech members since the estimate on the mediator is significant, and the mediator estimate is significant on performance -- Aguinis, Edwards & Bradley (2017) suggest that the direct effect on the outcome variable is actually not necessary to establish mediation. Further, the Sobel Test suggested a significant mediation effect (T-Stat,2.95; p, 0.003).

---

[7] Given potential dependency violations using network data in a linear regression model, we have replicated model 1 and 5 using the stochastic actor-oriented model in R Siena (Snjiders et al. 2010). This robustness check provided support for the hypothesis, where Sematech has positive significant estimate in the DOD period (est. 0.55; SE, 0.16; p<0.001) and a smaller but still positive and significant estimate in the post-DOD period (est. 0.41; SE, 0.14; p<0.001). Full results are available upon request.

**Table 2 – Mixed Effects Models: Mediation Analysis, DOD & Post-DOD Periods**

|              | DOD Period (1992-1995) | | | | Post-DOD Period (1996-1999) | | | |
| --- | --- | --- | --- | --- | --- | --- | --- | --- |
|              | E.Centrality | Performance | Performance | Performance | E.Centrality | Performance | Performance | Performance |
| Intercept    | 0.151*** | 50.49*** | 41.552*** | 43.06*** | 0.116*** | 76.256*** | 71.88*** | 72.744*** |
|              | (0.02) | (3.881) | (3.803) | (3.921) | (0.023) | (3.827) | (3.995) | (3.984) |
| Sematech     | 0.216*** | 19.452** |  | 10.039 | 0.263*** | 16.618** |  | 10.721 |
|              | (0.041) | (6.899) |  | (6.43) | (0.043) | (5.826) |  | (5.722) |
| S. Partners  | 0.036** | 1.087 |  | -0.099 | 0.068*** | -2.811 |  | -4.118 |
|              | (0.011) | (2.117) |  | (2.061) | (0.012) | (2.236) |  | (2.108) |
| E.Centrality |  |  | 48.954*** | 46.703*** |  |  | 23.317*** | 21.067** |
|              |  |  | (6.624) | (6.825) |  |  | (6.401) | (6.64) |
| Org.Size     | 0.00005*** | 0.004** | 0.003* | 0.002 | 0.00004*** | 0.002 | 0.002 | 0.001 |
|              | (0.00001) | (0.001) | (0.001) | (0.001) | (6.4E-06) | (0.0009) | (0.0008) | (0.0008) |
| US           | -0.013 | -3.346 | -2.491 | -4.364 | 0.006 | -6.941 | -9.217* | -9.184* |
|              | (0.022) | (4.455) | (3.896) | (4.071) | (0.026) | (4.441) | (4.402) | (4.381) |
| Year         | Fixed | Fixed | Fixed | Fixed | Fixed | Fixed | Fixed | Fixed |
| AIC          | -767.5 | 2521.2 | 2345.1 | 2333.7 | -668.9 | 2519.2 | 2331.8 | 2313.8 |
| Firm Est.    | 0.012 | 403.79 | 302.09 | 301.89 | 0.025 | 450.15 | 414.08 | 410.99 |
| Residual Est.| 0.007 | 124.63 | 116.91 | 116.61 | 0.008 | 118.91 | 104 | 100.29 |
| ICC          | 0.619 | 0.764 | 0.721 | 0.721 | 0.762 | 0.791 | 0.799 | 0.804 |
| Num.Orgs     | 196 | 196 | 161 | 161 | 231 | 231 | 192 | 192 |
| N            | 537 | 300 | 283 | 283 | 576 | 281 | 281 | 281 |

Figure Notes - p<.05*,p<.01**,P<.001***; standard errors in parentheses. The first four models are for a sub-set of the data between 1992-1995, while the second four models are between 1996-1999, reflecting the timeline in Figure 1. The order of the models represents the steps in mediation analysis. The sample sizes in the models are different because of the uneven data coverage across the variables for the unbalanced panel data. In addition, the coverage in data changes between the DOD and post-DOD periods.

Very similar patterns are observed in the post-DOD period from 1996 to 1999. However, there are important differences. Model 5 shows that Sematech's estimate on E.Centrality is larger in the post-DOD period. Yet Model 6 shows a reduced estimate of Sematech on Tech-Performance. Similarly Model 7 shows that E.Centrality has a lower estimate on Tech-Performance. Model 8 shows that Sematech retains a larger estimate despite including E.Centrality in the model, but with a p-value of only 0.063, indicating only a partial mediation effect (e.g. Baron & Kenny 1986). The Sobel Tests for Sematech (T-stat, 2.81; p, 0.005) and S.Partners (T-stat,2.77; p,0.006) showed the indirect effects are statistically different from zero.

These results provide support for hypothesis five, which specifies that estimates in hypothesis one through four are larger during the DOD/Maturity period and smaller in the post-DOD period.

To test whether the differences in effect size are significant between the DOD and post-DOD period, we employ an interaction term approach, where 1992-1995 are coded as 0 in the DOD period, and 1996-1999 are coded as 1 the post-DOD period. Table 3 shows the regression results. The results show that E.Centrality and Tech-performance are both generally higher in the post-DOD period with statistically significant estimates. Further, Sematech maintains a significant association with both E.Centrality and Tech-performance in the presence of interaction terms. However, in Model 1, the Sematech*post-DOD interaction term is not significant, while the S.Partners*post-DOD interaction terms is significant and positive, indicating that the allies of Sematech members had increases in network position during the post-DOD period. In model 2, the Sematech*post-DOD interaction term is significant and negative, indicating that Sematech members had higher relative Tech-performance during the DOD sponsorship period, consistent with hypothesis 5. Finally, in model 3 the E.Centrality*post-DOD interaction term is also negative and significant, indicating that the effect of network position on Tech-performance is stronger during the DOD period, which is also consistent with hypothesis 5.

**Table 3 – Mixed Models: Interaction Term Analysis (DOD/Post-DOD)**

|  | Interaction Term (DOD/Post-DOD) | | |
|---|---|---|---|
|  | E.Centrality | Performance | Performance |
| **Intercept** | 0.101*** | 49.085*** | 44.716*** |
|  | (0.019) | (3.292) | (3.276) |
| **Sematech** | 0.151*** | 17.042*** |  |
|  | (0.03) | (4.223) |  |
| **Post-DOD** | 0.033*** | 15.784*** | 40.658*** |
|  | (0.009) | (1.751) | (5.752) |
| **Sematech*Post-DOD** | 0.02 | -7.98* |  |
|  | (0.026) | (3.785) |  |
| **S. Partners** | 0.023 | 1.57 |  |
|  | (0.012) | (2.181) |  |
| **S.Partners*Post-DOD** | 0.077*** | -5.328 |  |
|  | (0.016) | (2.949) |  |
| **E.Centrality** |  |  | 14.276*** |
|  |  |  | (1.91) |
| **E.Centrality*Post-DOD** |  |  | -14.251** |
|  |  |  | (5.144) |
| **Org.Size** | 0.000*** | 0.002** | 0.002** |
|  | (0.000) | (0.001) | (0.001) |
| **US** | 0.000 | -5.377 | -6.28 |
|  | (0.021) | (3.943) | (3.708) |
| **AIC** | -1379.3 | 5114.5 | 4766 |
| **Firm Est.** | 0.017 | 429.18 | 340.82 |
| **Residual Est.** | 0.011 | 181.07 | 172.35 |
| **ICC** | 0.619 | 0.703 | 0.664 |
| **Num.Orgs** | 251 | 251 | 213 |
| **N** | 1113 | 600 | 564 |

Figure Notes - p<.05*,p<.01**,P<.001***; standard errors in parentheses. The models in the table are replications of model 1-3 in table 2 but have replaced the year variable for a DOD/post-DOD dummy variable used for interaction terms. The sample sizes are different in each model due to uneven data coverage for the unbalanced panel data.

Table 4 shows the robustness checks for the process model using two-way fixed effects models with year and organization fixed effects. The fixed effect model estimates the within rather than between organization effect and controls for omitted fixed variables. Due to low variability in Sematech membership and the outcome variable on separate four-year periods, these models are used on the full time period. The models in table 4 show a similar patter with the notable difference that Sematech remains significant in the final mediation model. This is consistent with combining the DOD and post-DOD periods in one analysis, since membership rapidly changed in 1996 due to entry of foreign firms in the consortium, only a partial mediation effect is suggested by the post-DOD period analysis in table 2. The Sobel Test is not significant (T-stat,1.82; p,0.067), failing to provide support for the mediation hypothesis.

**Table 4 – Fixed Effects Models: Mediation Analysis, Full Period**

|  | Full Period (1992-1999) | | | |
|---|---|---|---|---|
|  | E.Centrality | Performance | Performance | Performance |
| Intercept | 0.003 | 48.705*** | 46.036*** | 46.374*** |
|  | (0.100) | (5.563) | (6.952) | (6.885) |
| Sematech | 0.085** | 9.906** |  | 9.660** |
|  | (0.029) | (3.523) |  | (3.449) |
| S. Partners | 0.052*** | -1.524 |  | -1.732 |
|  | (0.009) | (1.535) |  | (1.517) |
| E.Centrality |  |  | 12.456* | 11.974* |
|  |  |  | (5.106) | (5.121) |
| Org.Size | 0.000** | -0.000 | -0.000 | -0.000 |
|  | (0.000) | (0.001) | (0.001) | (0.001) |
| Firm ID | Fixed | Fixed | Fixed | Fixed |
| Year | Fixed | Fixed | Fixed | Fixed |
| R2 | 0.835 | 0.864 | 0.865 | 0.868 |
| N | 1114 | 600 | 564 | 564 |

Figure Notes - p<.05*,p<.01**,P<.001***; standard errors in parentheses. Firm ID and Year are fixed in these models per the two-way fixed effects approach. The sample size for each model are different due to uneven data coverage for the unbalanced panel data.

## 6. Discussion

This study casts light on policy implementation through cross-sector inter-organizational networks. The empirical contribution is to reveal a previously invisible domain of cooperative activity. The results suggest policy effects on technological performance may be partially mediated through an organization's network position. By examining the dynamics of strategic cooperation in the formal R&D contract network of the semiconductor industry, this study suggests that network governance can play a part in the recovery and prosperity of the high-technology sector.

This study suggests how publicly sponsored consortia achieve public policy goals by facilitating cooperation between private-sector firms. Studies of organizational performance often avoid analyzing the relevance of government in explanations of strategic alliance behavior and firm performance, relying only firm-based resources (e.g. Barney 1991) and horizontal social relations (e.g. Burt 1997). Similarly, studies of science and technology policy tend to feature business and economics perspectives. Lambright (2008) suggested the need for analysis of S&T from a public policy and administration perspective. Scholars in public administration have analyzed the downstream process of contracting for complex technology (Brown, Potoski, Van Slyke, 2018). But what are governments to do if U.S. markets for these products fail, or if such vendors no longer have access to cutting edge platform technology, especially given the prospect that weapon systems or components must then be purchased from foreign nations and potential adversaries? In such cases government sponsored research consortia may prove to be a useful policy option.

Further, this study extends the network governance literature (Provan & Kenis 2008) to science and technology policy (e.g. Kash & Rycroft 1998; Laranja 2012; Wardenaar, de Jong, and Hessels 2014). The results suggest that government sponsored research consortia may be structured as network administrative organizations (NAOs) to achieve effectiveness in S&T networks. The results also suggest that public sponsorship through matching funds may provide needed support for building network capacity, while organizational exclusivity may enhance the network prominence of members. One advantage of studying Sematech as a mode of network governance (Provan & Kenis 2008) is that the theoretical tools used to analyze the private sector have been brought into public focus. Conversely, rather than conceptualizing the study only using public policy focused theories, the combination of policy frameworks with theory from business and economics broadens the perspective beyond power asymmetries and dependencies between private firms and central public agencies. Thus, the present study provides an integrated framework for S&T policy to explain how government action can both remedy market failures and catalyze innovation for increasing returns on investment.

*Limitations.* First, selection bias remains an issue in studies that do not use randomized selection for treatment and control groups. It may be possible that Sematech members self-selected into Sematech because they were already well performing firms. While there is likely some truth to this statement (in relative terms), it is important to bear in mind that Sematech was created to save *failing* U.S. business from foreign competition. Further, studies that estimate the effects of Sematech (Irwin and Klenow 1996; Link, Teece, and Finan 1996) show positive impacts on performance, but these do not examine the intermediary network effects. Further, the expectation is that Sematech is expected to have spillover effects on partners of Sematech members and on non-members, which further contaminates distinct group effects. Second, the chicken and egg problem between network position and performance persists, since firms that perform well will experience greater popularity in a competence-based collaboration network, i.e. preferential attachment in social networks (Wagner & Leydesdorff 2005). Third, it is unclear how the findings might generalize to other industries. For example, the network administrative organization model might not generalize effectively to industries where collaboration accounts for only a small or no portion of performance (cf. Provan & Lemaire 2012). Further, it is also unclear whether this approach would be appropriate in areas where strong principal-agent assumptions are necessary to maintain accountability (e.g. inherently government functions). Finally, the uneven data coverage across variables, the lack of network and performance data on years prior to the creation of Sematech and performance data during early implementation limits the analysis to mostly associational claims. Hence, while we believe the analysis supports the hypothesis, we avoid causal claims regarding the analysis.

**7. Conclusion**

Explanations for scientific and technological innovation often underestimate the role of government. However, science and technology are often best characterized as public/quasi-public goods prone to market failures absent government intervention. This study shows that governments might address market failures and enhance outcomes in science and technology by implementing policy through network governance. S&T policies can enhance technological performance for organizations engaged in strategic alliances by strengthening network capacity and effectiveness through government sponsored research consortia. Rather than only utilizing blunt policy instruments, such as direct subsidies or trade tariffs, the network administrative organization approach catalyzes local level cooperation, stimulates self-organization on dynamic technological landscapes, and enables synergies for collective advantage.

The findings may be generalized beyond the semiconductor industry to other sectors of interest to policymakers. For example, a comparison of the case of Sematech to emerging competition between the U.S. and China on artificial intelligence (AI) is striking. Rather than treating the issue as a "trade war" with trade tariffs and direct subsidies, implementation of industrial policy on AI could benefit from the network governance approach. Observers have suggested that Sematech is a model of industry-government cooperation (Hof, 2011), which might be applied to other areas. However, this study also points to the limitations of network governance because specific contextual and historical conditions appear to be critical for success. Cooperative policy instruments, such as the public-private consortium model, may be impossible to implement in sectors without a history of cooperative activity or in domains where inter-organizational networks are sparse or non-existent. Further, these types of arrangements may be inappropriate in sectors where strong principal-agent relations of hierarchical dependence are necessary to ensure accountability for inherently governmental functions. Finally, as Kingdon (1984) suggested, collective emergencies are often necessary to produce a cooperative atmosphere and a sense of urgency where leaders at odds might find common ground to pass legislation through windows of opportunity.